# Large depth of range Maxwellian-viewing SMV near-eye display based on a Pancharatnam-Berry optical element

Lin Wang,[1] Yan Li,[1,*] Shuxin Liu,[1] Yikai Su,[1] Di Wang,[2] and Qiong-Hua Wang[2]

[1] *Department of Electronic Engineering, Shanghai Jiao Tong University, Shanghai, 200240, China*
[2] *School of Instrumentation and Optoelectronic Engineering, Beihang University, Beijing, 100191, China*
*\*yan.li@sjtu.edu.cn*

**Abstract:** In order to overcome the accommodation and convergence (A-C) conflict that commonly causes visual fatigue in AR display, we propose a Maxwellian-viewing-super-multi-view (MV-SMV) near-eye display system based on a Pancharatnam-Berry optical element (PBOE). The PBOE, which is constituted with an array of high-efficiency polarization gratings, is implemented to direct different views to different directions simultaneously, constructing the 3D light field. Meanwhile, each view is like a Maxwellian view display that possesses a small viewpoint and a large depth of field (DOF). Hence, the MV-SMV display can display virtual images with correct accommodation depth cue within a large DOF. We implement a proof-of-concept MV-SMV display prototype with 3 × 1 and 3 × 2 viewpoints using a 1D PBOE and a 2D PBOE, respectively, and achieve a DOF of 4.37 diopters experimentally.

## 1. Introduction

Augmented reality (AR) [1-3], which augments the virtual information on the real-world environment, is an emerging technology that has potential applications in various fields such as healthcare, education and military. There are many AR wearable devices in the market, including Microsoft HoloLens, Magic Leap One and Google glasses. However, in most AR - displays, the virtual 3D images are generated by feeding the left and right eyes with two parallax images. In such a display, the visual axes of the two eyes are converged at the virtual 3D image depth, but the eye accommodation is fixed at the 2D image plane of the micro-display. The conflict between the accommodation depth cue and convergence depth cue is called accommodation and convergence conflict (A-C) conflict [4-9], and it will cause visual discomfort and fatigue after long-term use.

To address the A-C conflict issue in AR displays, different display methods have been proposed. The first one is multi-focal plane display [10-18], which generates 3D images by displaying the discrete 2D cross-section pictures of the 3D volume along the visual axis. The second method is vari-focal plane display [19], which dynamically adjusts the focal distance of a single-plane display to match the convergence depth of the eyes. The third one is Maxwellian-viewing (MV) display [20,21], where collimated image light is focused to a point at the pupil position by an eyepiece to provide an always-in focus image on the retina. Another promising approach is super-multi view (SMV) display [22-25]. In a SMV near-eye display, more than one parallax images are observed by a single eye, and therefore the light field of the multiple views can evoke correct focus adaption of the eye. However, this method usually suffers from a limited depth of field (DOF) [26].

In this paper, we propose a Maxwellian-viewing-super-multi-view (MV-SMV) near-eye display system based on a Pancharatnam-Berry optical element (PBOE) [27], to achieve both correct accommodation depth cue and a large DOF. The key component in our system, the PBOE, is composed of several regions of high-efficient gratings, that can deflect different views to different directions. The system also employs a holographic optical element (HOE) [28] which functions as an optical combiner and an eyepiece. In this MV-SMV display, the light of each view is produced from collimated laser light, and hence each view behaves like a Maxwellian-viewing display, providing always-in-focus images and a small eye box (or viewpoint). With multiple views, the light field of 3D objects can be faithfully reconstructed, while the DOF of the display is significantly improved due to the small eye box of each view. We have implemented the MV-SMV display system and achieved 3 × 1, and 3 × 2 viewpoints, respectively. Experimental result confirms that correct accommodation depth cue could be obtained within a large range from 20 cm to 160 cm.

## 2. Theory of depth of field

The working principle of a Maxwellian-viewing display is shown in Fig. 1 (a), where collimated image light is focused to a point by an eyepiece at the pupil position. It is as if each virtual image point only gives out one narrow beam in a certain direction, so in this pin-hole-like imaging system [29], no matter how the eye adjusts its focus, a clear image could always be formed on the retina, resulting in a large DOF. However, in a MV display, the eyebox is very small and the image could be easily missed out by the pupil.

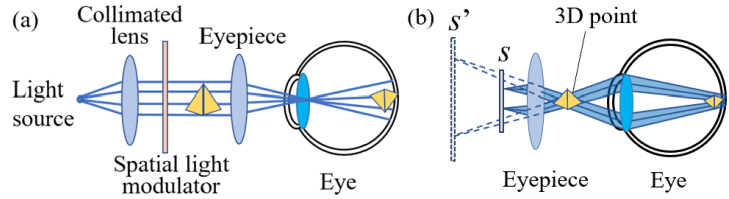

Fig. 1. Working principles of (a) a Maxwellian-viewing display and (b) a SMV display.

The working principle of a SMV display is shown in Fig. 1(b). Here, $s'$ is the virtual image of the screen (micro-display) $s$ formed by the display optics. So, if we ignore all the viewing optics for simplification, it could be considered as if the light beams of the parallax images are emitted by the pixels in the $s'$ plane in different directions. As multiple beams (from different views, respectively) enter the eye, the viewer is tempted to believe that 3D point is located at the intersection of the beams. To see the virtual 3D point clearly, the eye focus is so adjusted that the multiple beams are converged into one at the retina. So in this situation, the eye accommodation is consistent with the virtual 3D point instead of the 2D screen $s'$.

In most cases, the beams are divergent with finite beam sizes, so they will form an image spot instead of an image point on the retina. The more the 3D point deviates from s', the larger the size it is, and a more blurred image is formed on the retina. Since human eye has limited resolution, if the spot is smaller than the maximum acceptable size δ, which is about 15 μm [23], it could still be treated as a point of a clear image. So in order to generate clear images, we need to find out the range of the virtual point that could always produce a spot size smaller than δ. Such a range is the DOF of a SMV. If the virtual point is out of this range, no matter how the eye adjusts its focus, the virtual image is always blurry.

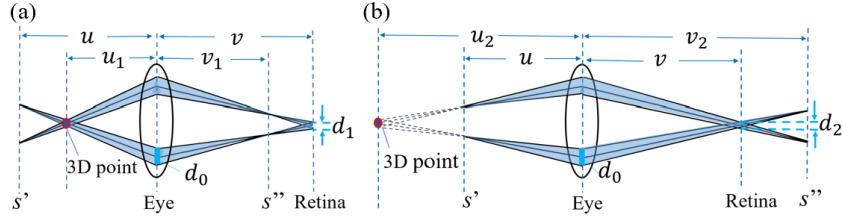

Fig. 2. Schematic diagrams of the focusing effect when the virtual image is (a) in front of and (b) behind the s' plane.

In the following, we will calculate the DOF of a general SMV display. Fig. 2(a) and (b) demonstrates the image formation of a virtual point in front of and behind the screen $s'$, respectively. In either case, according to the sets of conjugate planes, one could obtain the following relations:

$$\frac{1}{u_i} + \frac{1}{v} = \frac{1}{f_i} \tag{1}$$

$$\frac{1}{u} + \frac{1}{v_i} = \frac{1}{f_i} \tag{2}$$

where $f_i$ is the focal length of the eye when it is focused at the 3D image, $u_i$ is the distance from the 3D image to the human eye, $v$ is the distance from the human eye to the retina which could be considered as a constant, $d_0$ is the width of the light beam at the pupil position, $u$ is the distance from the screen image $s'$ to the human eye, and $v_i$ is the distance from the human eye to the $s''$ plane (the conjugate image of $s'$). Here, when virtual image is in front of the plane $s'$ as shown in Fig. 2(a), $i=1$. If it is behind $s'$ as shown in Fig. 2(b), $i=2$.

In addition, from the similar triangles in Fig. 2(a) and (b), one could obtain the following relations, respectively:

$$\frac{v - v_1}{v_1} = \frac{d_1}{d_0} \tag{3}$$

$$\frac{v_2 - v}{v_2} = \frac{d_2}{d_0} \tag{4}$$

Where $d_i$ ($i=1,2$) is the spot size of the image on the retina. When $d_1=d_2=\delta$, we can calculate the DOF in the form of diopters:

$$\Delta\phi = \frac{1}{u_1}\bigg|_{d_1=\delta} - \frac{1}{u_2}\bigg|_{d_2=\delta} = \frac{1}{v_1}\bigg|_{d_1=\delta} - \frac{1}{v_2}\bigg|_{d_2=\delta} = \frac{2\delta}{d_0 v} \tag{5}$$

From Eq. (5), one can see that DOF is inversely proportional to the width of the light beams at the pupil $d_0$. Therefore, a narrow beam width is favorable for a large DOF.

### 3. MV-SMV system

In the proposed MV-SMV system, for each view, the parallax image is produced from the collimated light source, so that it could be considered as a Maxwellian-viewing display. As multiple views projected from different directions are perceived by a single eye, SMV condition is satisfied and correct accommodation response could be evoked. Because of the narrow beam width in the MV-SMV display, the DOF is significantly enlarged. In the following, we will illustrate the working principle of the proposed MV-SMV display, introduce the two key components, the PBOE and the HOE, and demonstrate the experimental result as we have implemented a proof-of-concept prototype.

### 3.1 System configuration and working principle

The system configuration of the proposed MV-SMV display is shown in Fig. 3. Light coming from a 532 nm laser is expanded by a beam expander to illuminate a reflective amplitude-modulated liquid-crystal-on Silicon spatial light modulator (SLM). The image loaded on the SLM is an array of $N \times M$ parallax-view sub-images. Accordingly, the PBOE is also divided into $N \times M$ regions, which are basically high-efficient PB gratings with different periods and orientations. With precise alignment, the collimated light of a sub-image can only pass through the corresponding PB grating, and is deflected to a specific direction. The HOE (a Bragg volume optical element) and the refractive lens work together to converge different sub-image light into different viewpoints at the pupil position. When the intervals of the viewpoints are made sufficiently small, one eye could see more than one parallax sub-image, so that the 3D images can be reconstructed by the light field of multiple views, evoking correct accommodation response.

The two linear polarizers are appropriately arranged to achieve a high contrast for the SLM image. The right-handed circular polarizer (RCP) and a left-handed circular polarizer (LCP) are placed in front of and behind the PBOE, respectively, to ensure high diffraction efficiency [30].

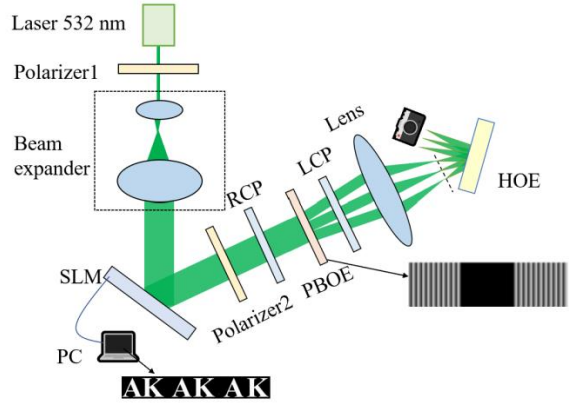

Fig. 3. Configuration of the MV-SMV system.

### 3.2 PBOE

PBOEs are thin diffractive devices with high polarization selectivity. They are essentially patterned half-wave plates with spatially varying optic-axis orientation. The working principle of PBOEs can be explained by Jones Matrix. The Jones vector of the incident circular polarization could be expressed as:

$$J_{\pm} = \frac{1}{\sqrt{2}} \begin{bmatrix} 1 \\ \pm j \end{bmatrix}, \qquad (6)$$

where $J_+$ and $J_-$ are for the left-handed circularly light and right-handed circularly polarized light, respectively. After passing through a small region of the PBOE where the optic axis of the half-wave plate is uniform, the Jones vector of the output light can be described as:

$$J'_\pm = R(-\phi)gW(\pi)gR(\phi)gJ_\pm = \frac{1}{\sqrt{2}}\begin{bmatrix} \cos 2\phi & \sin 2\phi \\ \sin 2\phi & -\cos 2\phi \end{bmatrix}\begin{bmatrix} 1 \\ \pm j \end{bmatrix} = \frac{1}{\sqrt{2}}\begin{bmatrix} 1 \\ mj \end{bmatrix}e^{\pm 2j\phi} \qquad (7)$$

where $\phi$, $W(\pi)$ and $R(\phi)$ are the azimuth angle of the optic axis, phase retardation of the half wave plate and rotation matrix, respectively [27,30]. As one can see here, the handedness of the circular polarization is inversed in addition to a phase change of $2\phi$. As the azimuthal angle $\phi$ changes continuously from 0 to $\pi$, the phase can be modulated from 0 to $2\pi$ continuously. This could avoid the discontinuous phase change and lead to high diffraction efficiency. If one could freely control the optic axis distribution of the PBOE, arbitrary phase profile could be generated.

Nowadays, with the advance of photoalignment technique, in-plane liquid crystal (LC) director distribution could be conveniently controlled, and various PB LC devices including gratings and lenses have been fabricated at low cost [31]. Fig. 4 (a) shows the LC directors in a continuous PB grating, which changes continuously and linearly from 0° to 180° in a period. And its phase distribution is shown in Fig. 4 (b). In our experiment, we fabricated 2 PBOEs (one has 3 × 1 regions and the other 3 × 2 regions) using a LC polymer material to achieve 3 × 1 and 3 × 2 viewpoints for the MV-SMV display, respectively. Each region of the PBOEs is a continuous PB grating whose period and orientation are appropriately designed to deflect light to a unique direction.

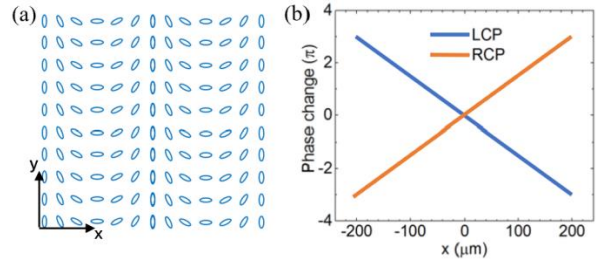

Fig. 4. (a) LC director distribution and (b) phase change of a continuous PB grating.

To fabricate the PBOE devices, first, the glass substrates were cleaned and then cured by a UV-Ozone for 15 mins. Then the photoalignment material, 0.4wt% Brilliant Yellow (BY) dissolved in dimethylformamide, was spin-coated onto the substrates at the speed of 500 rpm for 5 s and then 3000 rpm for 30 s [31]. Next, the sample was exposed to patterned polarization light field produced from a 488 nm laser with the density of 2 mW/cm$^2$ for 40 mins. Here, a non-interferometric setup [33] shown in Fig. 5 was employed to generate the patterned polarization field. It could generate an arbitrary alignment pattern by a single exposure. The SLM used in our experiment is a phase-type SLM (Holoeye, PLUTO-VIS) with a resolution of 1920 × 1080 and pixel size 8 μm. After exposure, a diluted liquid crystal reactive mesogen mixture (RMM) solution, consisting of 97wt% reactive mesogen RM257 and 3wt% photo initiator Irgacure 651 dissolved in toluene with a weight ratio of 1:3, was spin coated on the sample at the speed of 500 rpm for 5 s and then 3000 rpm for 30 s [32]. The same spin coating process was repeated once again to achieve the approximate thickness of a LC half wave plate. Immediately after that, the RMM was cured by 365 nm UV light to form a polymer film.

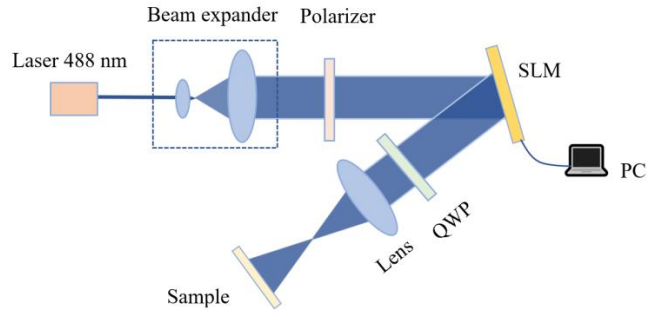

Fig. 5. Optical setup of PBOE exposure.

Fig. 6 (a) shows the microscopic pictures of the different regions of the 1D PBOE (3 × 1 regions). The middle region has uniform alignment, so that light can directly pass through without any deflection. The right and left regions have the same gratings period (48 μm) in the x direction, but complementary phase profiles. The expected diffraction angles of the two regions are 0.64° and −0.64°, respectively, according to the grating equation $d\sin\theta = \lambda$ ($\lambda = 532\ nm$). The diffraction patterns of the two regions with circularly polarized incidence are shown in Fig. 6 (c). The diffraction efficiencies of the right and left regions are 93.2% and 88.8%, respectively. Here, the diffraction efficiency is defined as the ratio of the 1$^{st}$ order diffracted intensity to the total light intensity collected after the PB grating.

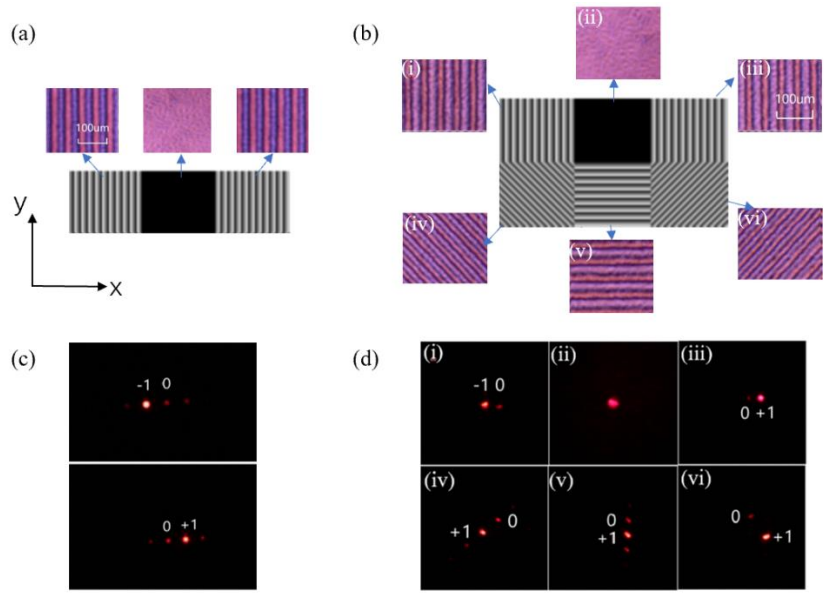

Fig. 6. (a) Microscopic pictures of different regions in the 1D PBOE and (b) in the 2D PBOE. (c) Diffraction patterns of the right (top) and left regions (bottom) of the 1D PBOE. (d) Diffraction patterns of different regions of the 2D PBOE.

Fig. 6 (b) shows the microscopic pictures of the different regions in the 2D PBOE. The LC alignment of the first row is exactly the same as the 1D PBOE. The orientations of the three gratings in the second row are 45°, 90° and -45°, with respect to x direction in the x-y plane, respectively, and their grating periods are 34 μm, 48 μm and 34 μm. The diffraction efficiencies of the regions (i), (iii), (iv), (v) and (vi) are 93.1%, 89.3%, 85.1%, 89.5%, 82.6%, respectively, and the diffraction patterns are shown in Fig. 6 (d).

### 3.3 HOE

HOEs are optical elements fabricated by exposing holographic recording materials to interfered laser light. HOEs usually exhibit high wavelength and angle selectivity due to their Bragg structure nature. For light satisfying Bragg condition, a HOE can realize a certain optical function like a grating, a lens or a mirror [34,35], with high diffraction efficiency. For light dissatisfying Bragg condition, however, it is just like a transparent window. Because of these characteristics, HOEs have been employed as optical combiners in AR systems [36,37].

In our experiment, we fabricated a HOE to function as a mirror lens in the optical system. Fig. 7(a) illustrates the recording process of the HOE. The HOE sample is placed at the position where a spherical wave and a plane wave, coming from opposite sides, interfere. The angle $\alpha$ between reference beam and the HOE sample plane is approximately 45°, and the spherical wave impinges on the sample normally. The HOE material used in the experiment is a commercial holographic film (Litiholo C-RT20), and the required exposure energy is ~30 mJ/cm$^2$ at the wavelength of 532 nm [38]. In our experiment, we exposed the HOE for half an hour. The density of the object beam and reference beam are both approximately 50 mW/cm$^2$. The focal length of the lens used for generating the object beam is 100 mm.

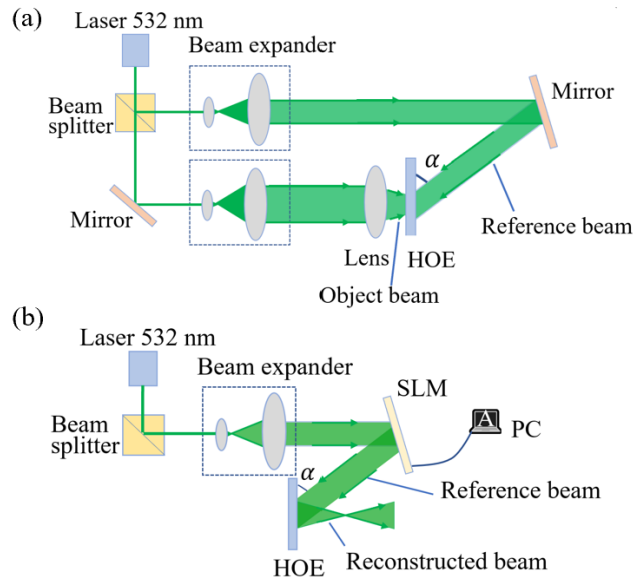

Fig. 7. (a) Recording process and (b) reconstruction process of the HOE.

The measured diffraction efficiency of the HOE lens is ~60%, and its focal length is ~70 mm. To evaluate its imaging quality, collimated light modulated with a letter "A" was projected to the HOE as shown in Fig. 7 (b). Fig. 8 (a) shows the focus pattern received on a screen, which indicates good focusing ability. And Fig. 8 (b) shows the projected image when the screen is placed at a farther distance.

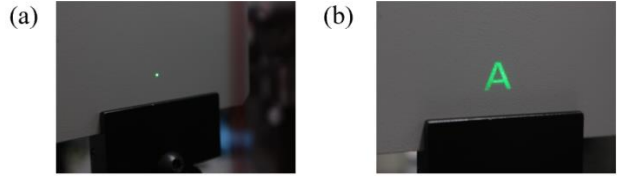

Fig. 8. (a) Focus pattern of the HOE lens and (b) projected image by the HOE.

### 3.4 Prototype implementation and result

We implemented a proof-of-concept MV-SMV display prototype according to Fig. 3, and achieved 3 × 1 and 3 × 2 viewpoints using the 1D and 2D PBOEs, respectively. The SLM used in the experiment is an amplitude-type SLM (Holoeye, LC-R-1080), whose the resolution is 1920 × 1080 and the pixel size is 8 $\mu m$. The RCP in front of the PBOE is to ensure circular polarized incidence. Although the diffraction efficiencies of the PB gratings are high, we still placed a LCP behind the PBOE to eliminate the zero-order light whose polarization was unchanged. The focal length of the refractive lens is 60 mm.

The refractive lens and the HOE work together to converge the collimated light beams into viewpoints at the pupil position. Fig. 9 (a) shows the photo of the viewpoints received on a screen at the exist pupil of the system with 3 × 1 viewpoints. The interval of the adjacent viewpoints $d$ is ~ 0.8 mm and the size of each viewpoint $d_0$ is ~0.20 mm. Since the pupil diameter of a human eye is usually 4-8 mm, with such small viewpoint intervals the system can satisfy the SMV condition. The viewpoint interval $d$ is determined by the deflection angle of the PBOE $\theta$ and the focal length of the lens system (comprised of the refractive and HOE lenses) $f$ as $d = f\tan\theta$, based on the simplified configuration shown in Fig. 10.

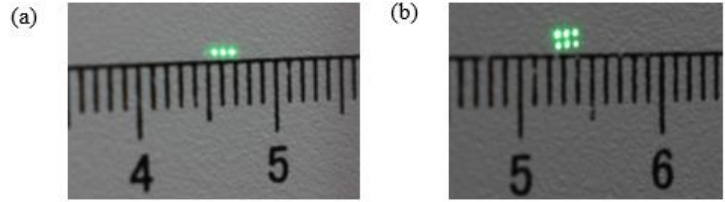

Fig. 9. Photos of the viewpoints at the pupil position: (a) 3 × 1 viewpoints and (b) 3 × 2 viewpoints.

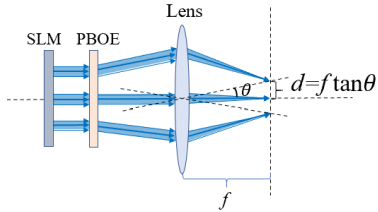

Fig. 10. Calculation of viewpoint intervals in a simplified configuration. Lens -- the equivalent lens of the refractive lens and HOE lens.

For each view of the MV-SMV, it is a like Maxwellian-viewing display, and should generate always-in-focus images within a large DOF. To confirm this, we only displayed the middle view with a letter "A" while showing nothing in the other views. A camera was placed near the exist pupil to took photos. As the camera focus was adjusted from near distance to far distance, the captured images were always sharp though their sizes varied. Fig. 11 (a) and (b) shows the pictures taken when the camera was focused at 20 cm and 200 cm, respectively, for example.

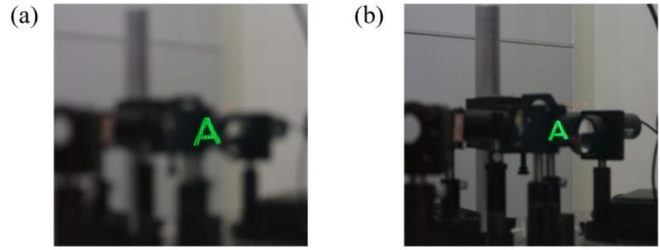

Fig. 11. Captured images from the middle view when camera was focused at (a) 20 cm and (b) 200 cm.

When three parallax images (letters "A" and "K" with different intervals) were displayed, virtual 3D images could be generated. Here, the virtual image letter "A" was generated at the depth of 160 cm, while the letter "K" was at 20 cm. Fig. 12 (a) and (b) shows the photos of the virtual 3D images augmented on the real world, when the camera was focused at 20 cm and 160 cm, respectively. We can see that when one of the letters is clear in focus, the other is blurry. The virtual images exhibit the same in and out-of-focus effects as the real objects (an aperture at a near distance and a notebook at a far distance). Therefore, this MV-SMV display could indeed provide correct accommodation depth cue for virtual 3D images. Moreover, within the large range of depth from 20 cm to 160 cm (or from 5 diopters to 0.63 diopters in the form of diopters), we could always observe clear virtual images, indicating a large DOF of at least 4.37 diopters.

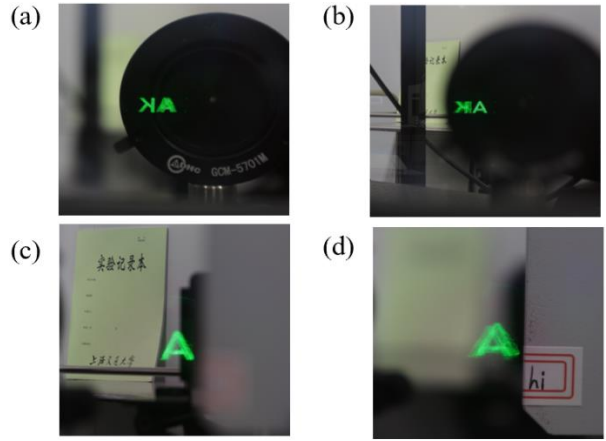

Fig. 12. Captured images of the MV-SMV display with 3 × 1 viewpoints when camera was focused at 20 cm and (b) 160 cm. Captured images of the MV-SMV display with 3 × 2 viewpoints when camera was focused at (c) the notebook and (d) the 'hi' label.

We further implemented the system to achieve 3 × 2 viewpoints using the 2D PBOE. Six parallax images of letter "A" were projected from different regions of the SLM, respectively. Similarly, the light beams were converged to different viewpoints as shown in Fig. 9 (b). When camera was focused at the notebook which is 160 cm away, the letter "A" was also clear in focus. When the camera was focused at a "hi" label, the virtual image "A" is blurred just as the notebook. In this experiment, with 3 × 2 viewpoints, both horizontal and vertical parallaxes were realized, and the full parallax could provide more natural 3D effect.

## 4. Conclusion

In this study, we developed and demonstrated a MV-SMV near-eye display system based on a PBOE. The collimated parallax images are deflected by the PBOE into different directions, and are then converged into viewpoints by the lenses. We implemented the MV-SMV AR display with 3 × 1 and 3 × 2 viewpoints, using a 1D PBOE and 2D PBOE, respectively. The HOE serves as both an eyepiece and a combiner in the AR display system. The spot size of each viewpoint is about 0.2 mm, and the interval of the adjacent viewpoints is about 0.8 mm. The small interval ensures the satisfaction of SMV condition that more than one viewpoint should be observed by a human eye simultaneously. Thanks to the small size of the viewpoints, the system can display 3D images with correct accommodation depth cue within a large DOF of at least 4.37 diopters. We believe the proposed MV-SMV display holds great promise for future AR display applications.

**Funding.**

**Disclosure.** The authors declare no conflicts of interest.